\documentclass[aps,pra,groupedaddress,superscriptaddress,twocolumn,10pt,longbibliography]{revtex4-2}
\pdfoutput=1 
\usepackage[utf8]{inputenc}
\usepackage{hyperref}
\usepackage{graphicx}
\usepackage{orcidlink}
\usepackage{booktabs}
\usepackage{tabularx}
\usepackage{cancel}

\newcommand{\cmtout}[1]{}

\usepackage{siunitx} 
\newcommand{\omatt}[1]{}

\usepackage{soul} 
\usepackage[normalem]{ulem}
\usepackage{amssymb}
\usepackage{physics}
\usepackage{hyperref}
\DeclareSIUnit\bohr{\text{\ensuremath{a_\textup{0}}}}

\begin{document}

\title{Observation of spin singlet butterfly Rydberg molecules in an ultracold atomic Rb gas}

\author{Markus Exner\orcidlink{0009-0005-8290-7371}}
\affiliation{Department of Physics and Research Center OPTIMAS, Rheinland-Pfälzische Technische Universität Kaiserslautern-Landau, 67663 Kaiserslautern, Germany}

\author{Rohan Srikumar\orcidlink{0000-0003-0303-1331}}
\affiliation{Zentrum für Optische Quantentechnologien, Universität Hamburg, Luruper Chaussee 149, 22761 Hamburg, Germany}

\author{Richard Blättner\orcidlink{0009-0004-4667-821X}}
\affiliation{Department of Physics and Research Center OPTIMAS, Rheinland-Pfälzische Technische Universität Kaiserslautern-Landau, 67663 Kaiserslautern, Germany}

\author{Peter Schmelcher\orcidlink{0000-0002-2637-0937}}
\affiliation{Zentrum für Optische Quantentechnologien, Universität Hamburg, Luruper Chaussee 149, 22761 Hamburg, Germany}
\affiliation{The Hamburg Centre for Ultrafast Imaging, Universität Hamburg, Luruper Chaussee 149, 22761 Hamburg, Germany}

\author{H. R.  Sadeghpour\orcidlink{0000-0001-5707-8675}}
\affiliation{ITAMP, Center for Astrophysics $|$ Harvard \& Smithsonian, Cambridge, MA 02138 USA}

\author{Matthew T. Eiles\orcidlink{0000-0002-0569-7551}}
\affiliation{Max Planck Institute for the Physics of Complex Systems,  Nöthnitzer Str. 38, 01187 Dresden, Germany}

\author{Herwig Ott\orcidlink{0000-0002-3155-2719}}
\thanks{Corresponding author: ott@physik.uni-kl.de}
\affiliation{Department of Physics and Research Center OPTIMAS, Rheinland-Pfälzische Technische Universität Kaiserslautern-Landau, 67663 Kaiserslautern, Germany}

\date{\today}

\begin{abstract}
We report the observation of spin-singlet ultra-long range Rydberg butterfly molecules consisting of a ground-state atom bound to a Rydberg atom by $P$-wave scattering of $^{87}$Rb Rydberg electrons from $^{87}$Rb(5s) atoms. A three-photon excitation scheme enables the photoassociation of these molecules by weakly admixing Rb($18f_{7/2}$) states. The measured binding energies, kilo-Debye permanent electric dipole moments, and lifetimes are in excellent agreement with theory. Two long-lived vibrational levels, red detuned from the Rb($18f_{7/2}$) threshold, are observed. This experiment is a foundational step in the production of ultra-cold anions and heavy Rydberg ion-pair systems.

\end{abstract}

\maketitle

The zoology of ultralong-range Rydberg molecules (ULRMs)  categorizes its subjects by the orbital angular momentum ($\ell$) of the Rydberg electron \cite{Shaffer2018,Eiles_2019,Hummel2020,Dunning_2024}.
While a low-$\ell$ Rydberg state in possession of a sizable quantum defect 
remains approximately spherically symmetric as it binds to a ground-state atom \cite{Bendkowsky2009,killian2015,Deiglmayer2021}, molecules with high-$\ell$ character have highly asymmetric electronic states \cite{Greene2000, Hamilton_2002, Khuskivadze_2002,GiannakeasIon2020}.
These molecules, known colloquially as trilobites and butterflies, possess large permanent electric dipole moments (PEDM) and have been observed in ultracold gases of alkali atoms 
in predominantly spin-triplet configurations \cite{Sadeghpour_2011,Booth_2015,Niederprum2016,Althoen2023,Exner2024}. 
Spin-singlet configurations lead to much shallower, or even repulsive, molecular potentials, and thus to date have only played an indirect role through the mixing of singlet and triplet states via the hyperfine structure of the ground-state atom \cite{Deiglmayer2015,Kleinbach_2017, bottcher2016observation}.

\begin{figure}[hb!]
\includegraphics[scale=0.98]{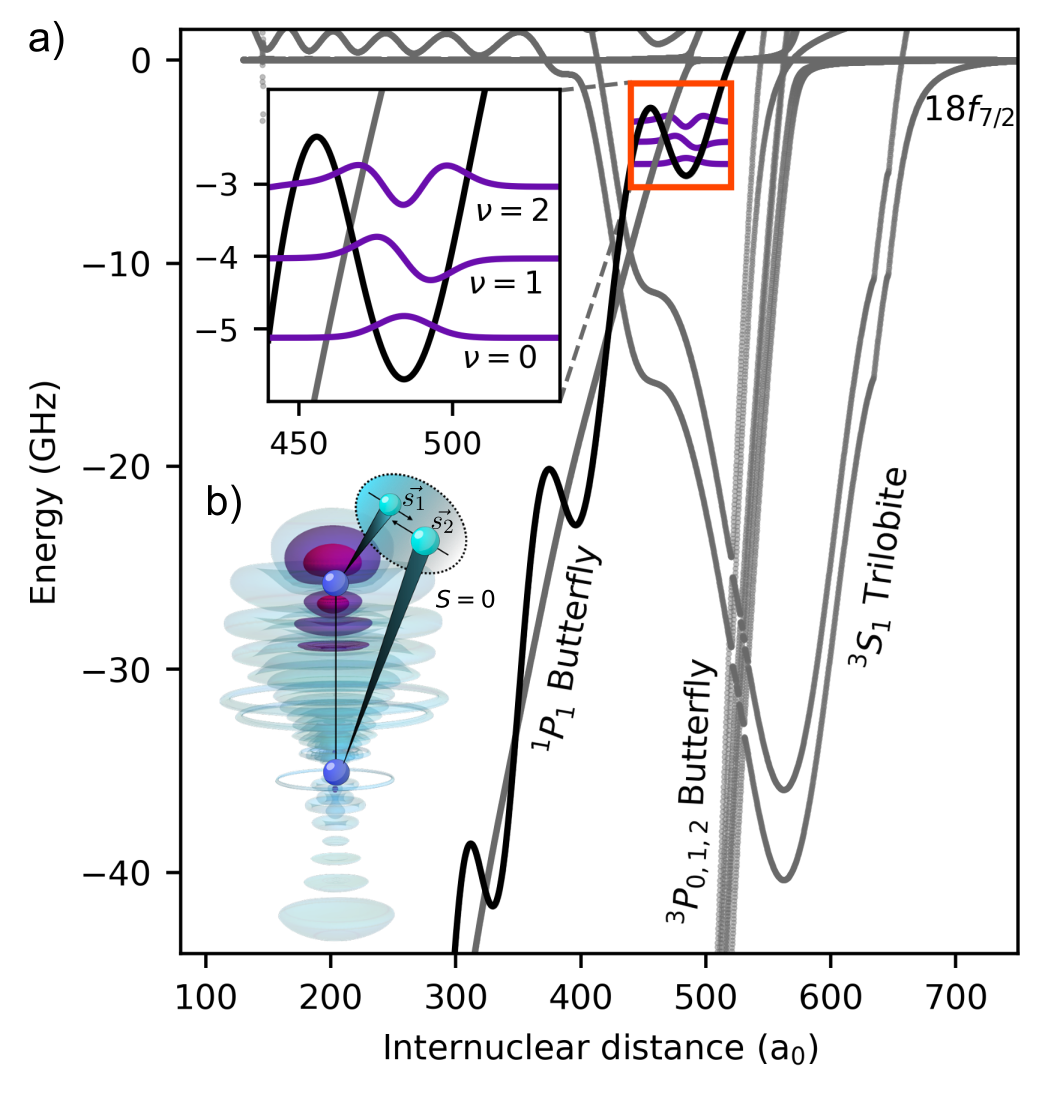}
\caption{ a) Born-Oppenheimer potential energy curves for the $n=18$, $\Omega = 3/2$ $^{87}$Rb$_2$ ULRM. The black curve highlights the potential curve with $M_L \approx 0$ whose oscillations are deep enough to support vibrational states. The red box marks the first butterfly well, magnified in the inset with the calculated vibrational states. b) Electron density distribution of the spin-singlet butterfly molecule. 
}
\label{fig:potential}
\end{figure}

Ultracold neutral plasmas have been created \cite{Killian2007,Killianplasma1999,Zelenerplasma2024,Fletcherplasma2006}, sparking the exploration of exotic phases such as the frozen-Rydberg gas \cite{Gallagher2004,Gallagher2000,Gallagher2005} and Coulomb crystal \cite{pohl2004} in ultracold gases.
Ordinary two-component ultracold plasmas, created by ionizing ultracold atoms, suffer from a variety of heating mechanisms mainly because one component (electron) is always much lighter and faster, thus leading to heating and spontaneous three-body recombination and Rydberg formation \cite{vrinceanu2008}. 
In 2013, it was proposed to create an ultracold two-component plasma in a MOT utilizing charge transfer to form ion pairs out of Feshbach molecules \cite{kirrander2013, Markson_2016}. 
Recently, a scheme utilizing the spin-singlet ($^1P_1$) butterfly URLM as a central step to form heavy Rydberg states (HRS) has been developed  \cite{Hummel_2020}.

Heavy Rydberg states, the molecular analogues of Rydberg atoms, can dissociate into positive and negative atomic ions, thus providing a pathway towards production of ultracold anions avoiding known challenges posed by the lack of laser-cooling transitions in typical anions or difficulties with sympathetic cooling \cite{Walter2014,Tang2019}.
Therefore, the production of singlet URLM states is an important first step towards the creation of strongly-coupled two-component plasmas and ultracold anions. 

In this work, we report the observation of a $^{87}$Rb butterfly long-range Rydberg molecule in the spin-singlet ($^1P_1$) configuration.
We use a three-photon excitation scheme to photoassociate the $n$=18 singlet butterfly molecule via the small $f$-state admixture in its electronic state. 
Two long-lived vibrational states bound within a well lying about 5GHz below the 18$f_{7/2}$ atomic resonance are detected.  The measured binding energies, dipole moments and lifetimes of these levels are in excellent agreement with theoretical predictions, confirming the $^1P_1$ characterization of the molecule.

The binding mechanism of an ultralong-range Rydberg molecule is mediated by the scattering of the Rydberg electron from the ground-state atom a distance of $R$. 
Scattering via $S$ and $P$ partial waves dominates because of the Rydberg electron's low kinetic energy. 
The phase shifts in these partial waves are sensitive to both the spin configuration (singlet and triplet) and, for heavy atoms such as Rb and Cs, the fine-structure in the $P$-wave channel. 
We include the six dominant scattering channels $^{2\mathrm{S} + 1}L_J$: $^1S_0$, $^1P_1$, $^3S_1$, and $^3P_{0,1,2}$. 
The resulting high$-\ell$ molecular states are largely decoupled from one another.

\begin{figure}[t]
\centering
\includegraphics[scale=1]{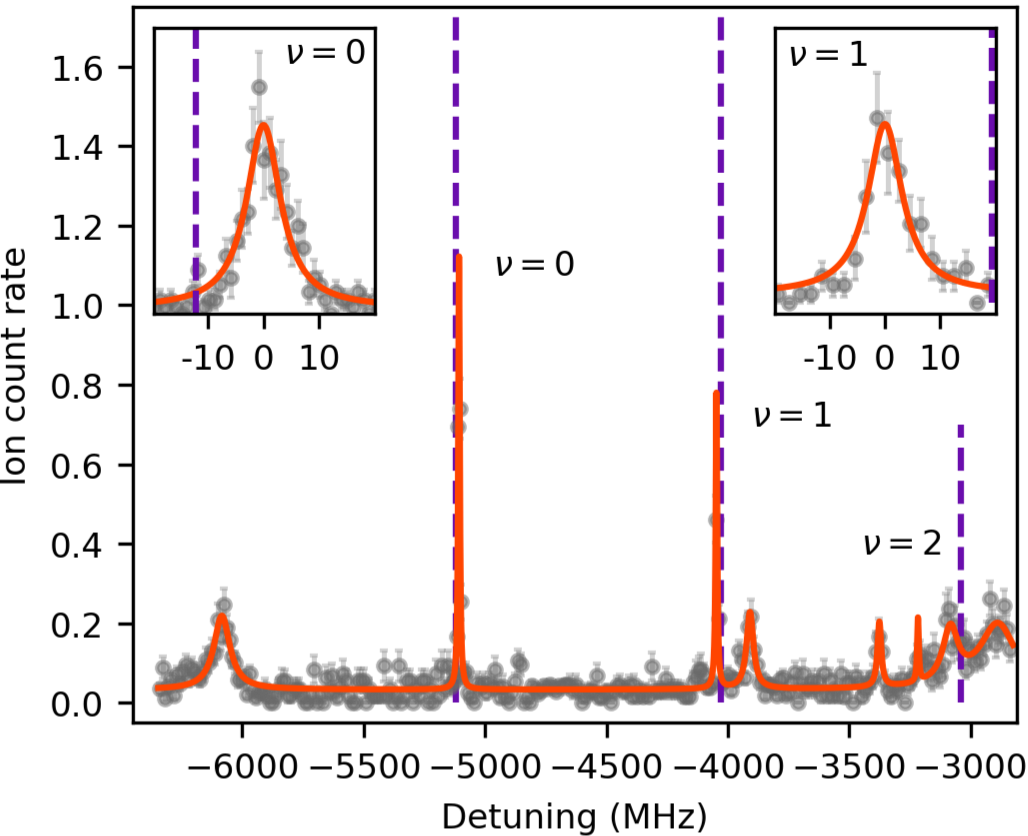}
\caption{Ion Rb$^+$ signal (binned) as a function of frequency detuning relative to the Rb(18$f_{7/2}$) threshold. The first two molecular singlet butterfly resonances are magnified in the two insets. The calculated eigenenergies of all three vibrational states are shown as dashed vertical lines.  
Please note that the x-axes (detuning) of the insets take the corresponding molecular resonances as their energy zeros,  not the atomic resonance
(like in the main figure).
}
\label{fig:spectrum}
\end{figure}

We obtain the adiabatic potential energy curves from the $R$-dependent eigenvalues of the electronic Hamiltonian including the full spin-dependent structure of both Rb atoms \cite{Eiles_2017} and computed using a Green's function treatment \cite{eiles2023}. 
Fig.~\ref{fig:potential}a) shows the adiabatic PECs for $\Omega = 3/2$, where $\Omega$ is the projection of the total angular momentum onto the internuclear axis. The dissociation threshold corresponds to the ($\ket{18f_{7/2}}$ +  $\ket{5s_{1/2}, F=1}$) pair state. 
The $^1P_1$ scattering produces the oscillating ``stairwell" potential highlighted in black as well as  the non-oscillating potential slicing through the stairwell.  
Although the projection of the orbital angular momentum $M_L$, is not separately conserved, these two potentials are predominantly of $M_L = 0$ and $|M_L| = 1$
character, respectively, due to the very weak spin-orbit coupling \cite{Deiss2020,Rodriguez2024}. Unlike the $^3S_1$ and $^3P_J$ phase shifts, the $^1P_1$ phase is a monotonic and concave function of the electron momentum $k$. Hence the PECs exhibits local minima only due to the oscillatory Rydberg electronic density.

The first well in this stairwell potential supports three bound vibrational states (see inset of Fig. \ref{fig:potential}a)). Their energetic proximity to the $18f_{7/2}$ resonance, and the $f$-state electronic admixture they possess make it possible to access these vibrational states via three-photon excitation.
To observe these states, we prepare $^{87}$Rb atoms in a crossed dipole trap  operating at $\lambda = \SI{1064}{nm}$ with a temperature of \SI{40}{\micro K}, a diameter of \SI{40}{\micro m} and a peak density of \SI{4 e13}{cm^{-3}}. Initially, the atoms are in the F=1 ground state and photoassociation in the molecular state is achieved by a three-photon excitation scheme (5$s_{1/2}\to $5$p_{3/2} \to $5$d_{5/2} \to$$18f_{7/2}$) at 780 nm, 776 nm and 1308 nm, where the first two lasers are blue detuned to the intermediate states. After excitation, the Rydberg atoms are ionized by a CO$_2$ laser and the ions are detected in a reaction microscope. The experimental sequence consists of 1100 excitation and ionization pulses with a duration of $t_\mathrm{exc}= \SI{3}{\micro s} $. During excitation, the dipole trap is turned off to avoid ionization from the $5d_{5/2}$ state.   Due to the high field sensitivity of the molecules, the electric field is switched off during photoassociation with a residual field $E_\mathrm{residual} \approx 1 \, \mathrm{mV/cm}$.

Figure \ref{fig:spectrum} shows the spectrum with energies measured relative to the $18 f_{7/2}$ atomic resonance. The solid red curve shows a fit of eight Lorentzian peaks. The spectrum clearly exhibits two narrow main peaks, with a full width at half maximum (FWHM) of about \SI{7}{MHz}. These resonances can be assigned to the ground and first-excited vibrational states in the first well of the stairwell potential, below the $18 f_{7/2}$ atomic resonance (Fig. \ref{fig:potential}a)).  Compared to the pure trilobite resonances observed in Refs. \cite{Althoen2023,Exner2024}, the FWHM is the same, but the signal strength is significantly lower. We attribute the lower signal strength to the smaller molecular bond length, which reduces the Franck-Condon factor due to the reduced probability of finding two atoms at the correct separation in the ultracold gas.

The spectrum also contains several peaks with substantially lower signal strength. These resonances have a high FWHM which is more than twice as large as that of the two larger peaks. One of these peaks fits well with the second excited vibrational state ($\nu=2$) at a detuning of about \SI{-3}{GHz}. The lower signal strength can be attributed to a shorter lifetime due to molecular decay; this state has a much higher tunneling rate through the potential barrier separating it from the short-range region where $\ell$-changing predissociation collisions become dominant. 
The linewidth of this state \footnote{We use Siegert pseudostates to compute the resonance positions and widths.}\cite{durst2024} is $\sim\SI{10}{MHz}$, which is much larger than that of the first two vibrational states ($\sim$kHz).

The origin of the other low signal peaks is not {well} understood.
While highly excited trilobite states (stabilized through non-adiabatic effects \cite{srikumar2025,eiles2024}) can exist for $n=18$ in this energy range, the computed level spacing of these states is substantially smaller than that seen between the observed resonances. 
There are additional molecular potential curves associated with higher partial wave $(L\ge 2)$ scattering present in this energy range at still smaller internuclear distances \cite{GiannakeasIon2020}, and thus a possible candidate for these features could be "dragonfly" Rydberg molecules. A quantitative calculation to verify this requires the extension of the spin-coupled Green's function treatment to include $L=2$ partial waves, and thus further characterization of these states remains the subject of future work.

 \begin{table}[t]
    \centering
    \caption{Binding energies, full width at half maximum (FWHM), dipole moments and lifetimes of the butterfly molecules }
    \label{Tab}
    \begin{tabular*}{\columnwidth}{@{\extracolsep{\fill}}lccc}
    \toprule
         & $\nu = 0$  & $\nu =1$   & $18f_{7/2}$  \\
        \midrule
        $E_\mathrm{b,Exp}$ (MHz) & $5107 \pm 2$ & $4047 \pm 2$ & {} \\[0.6ex]
        $E_\mathrm{b,Theo}$ (MHz) & $5119 $ & $4027 $ &  {}\\[0.6ex]
        FWHM (MHz) & $7.2 \pm 0.4$ & $7.1 \pm 0.9$ &  {}\\[0.6ex]
        d$_{\mathrm{Exp}}$  (Debye) & $851 \pm 104$ & $857 \pm 104$ & {}  \\[0.6ex]
        d$_{\mathrm{Theo}}$ (Debye) & \num{840} & \num{823} & {}  \\[0.6ex]
        Lifetime (\si{\micro\second}) & $6.31 \pm 0.67$ & $1.79 \pm 0.19$ & $4.15 \pm 0.06$ \\[0.6ex]
        \bottomrule
    \end{tabular*}  
\end{table}

Recent works have elaborated on the extraction of $^3S_1$ phase shifts \cite{Exner2024} and $^3P_J$ resonance positions \cite{Engel_2019} using spectral analysis of the ULRM. 
In the present calculation, we use the energy-dependent $^3S_1$ and $^3P_J$ phase shifts fitted to trilobite spectra in Ref.~\cite{Exner2024}, as they accurately describe the $S=1$ trilobite and butterfly curves.
The $^1P_1$ phase shift for Rb were taken from Ref.~\cite{Eiles_2018}, which is in good agreement with that calculated by Ref.~\cite {Chibisov_2002}. 
The $^1P_1$ phase shift was modified by an energy-independent factor of 1.02
, which shifts the well depth by around -435 MHz and leads to the excellent match between theoretically predicted and observed molecular resonances seen in Fig.~\ref{fig:spectrum}. 
A perturbative estimate shows that contributions from $L=2$ scattering could have an effect on the level of several tens of MHz.

\begin{figure}[t]
\centering
\includegraphics[scale=1]{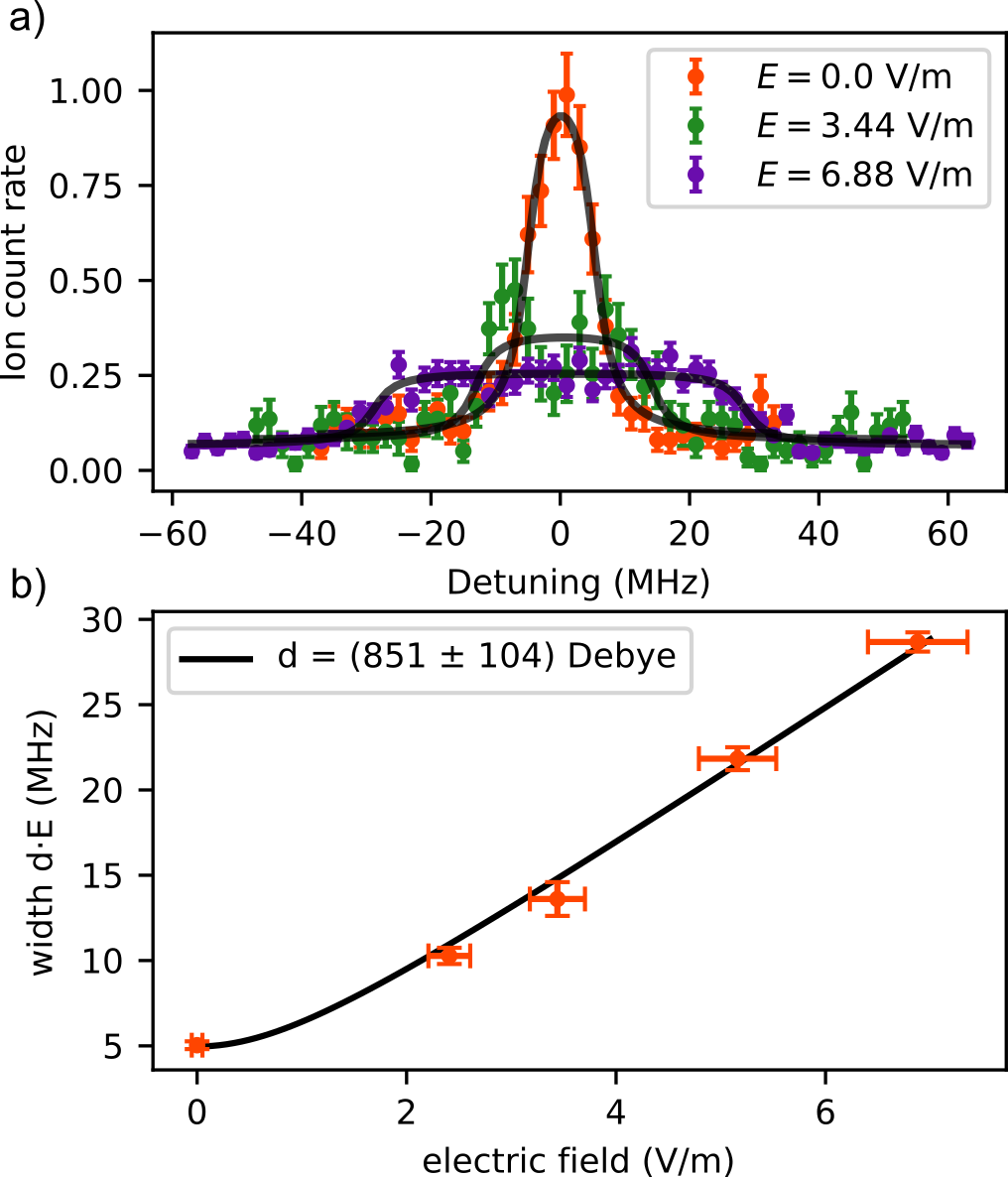}
\caption{Measurement of the dipole moment for the $\nu=0$ butterfly state. a) Spectrum of the molecular resonance for different electric fields fitted with a convolution of a Lorentzian and two step functions of width $2dE$. b) Width $dE$ as a function of the electric field fitted by $d\sqrt{E^2+E_0^2}$ to obtain the dipole moment of \mbox{$d=851 \pm 104 \, \mathrm{Debye}$}. }
\label{fig:dipole}
\end{figure}

\begin{figure}[t]
\centering
\includegraphics[scale=0.97]{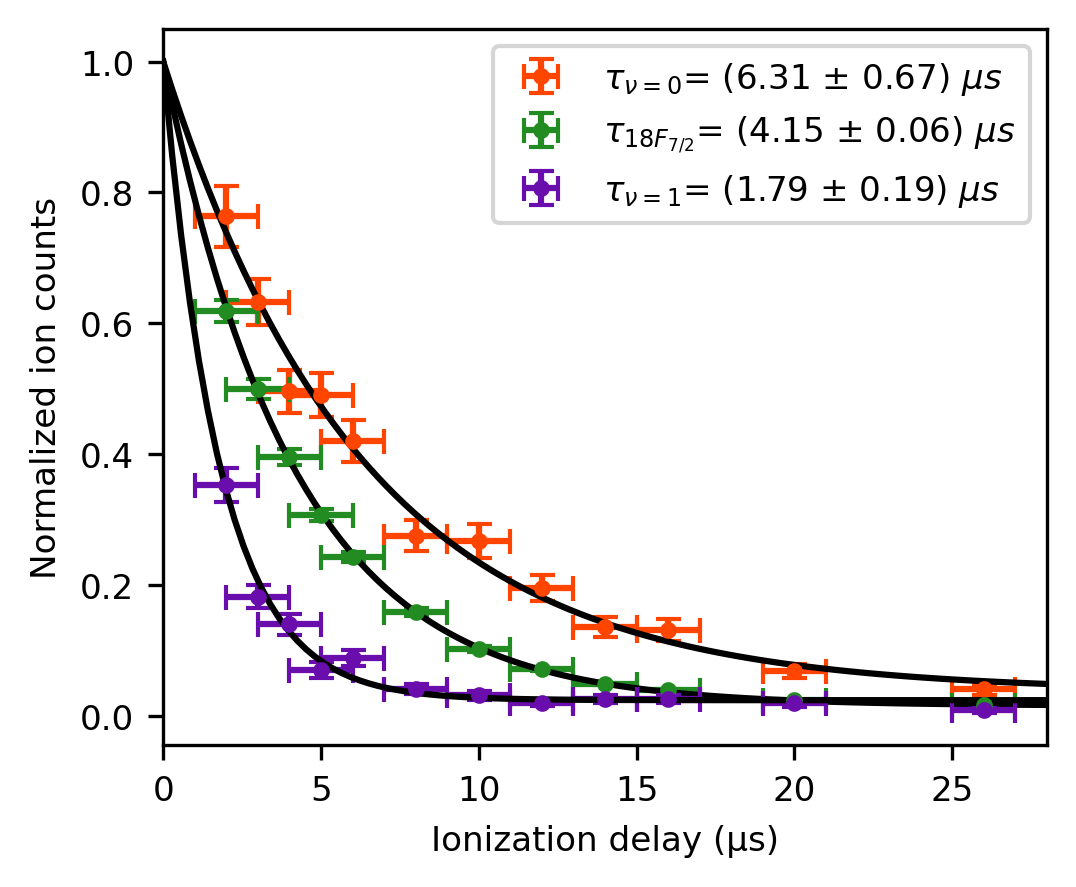}
\caption{Normalized ion counts as a function of the delay between excitation and ionization for the two molecular and the atomic $18f_{7/2}$ state. The high $\ell$-admixture leads to an increased lifetime of the molecule. The shorter lifetime for the $\nu=1$ state is likely due to the avoided crossings with $M_L=\pm 1$ potentials.   }
\label{fig:lifetime}
\end{figure}

We next characterize the observed two main peaks by their electric dipole moments and lifetimes to confirm that they correspond to singlet butterfly molecules.
Since the butterfly potential originates from the degenerate hydrogen manifold and couples high angular momentum states with different parities,  these high-$\ell$ ULRM exhibit linear Stark shifts. To determine the dipole moment, the broadening of the molecular resonance is measured for different electric fields and fitted with a convolution of a Lorentzian and two step functions of width $2dE$ (see Fig. \ref{fig:dipole}a). Fig. \ref{fig:dipole}b) shows the fitted $dE$ as a function of the electric field, which leads to an almost identical dipole moment of  \SI{851}{Debye} and \SI{857}{Debye}. The theoretical dipole moment was calculated using electronic states obtained by diagonalizing the electronic  Hamiltonian in a truncated basis of Rydberg states after benchmarking the basis size using the Green's function results \cite{Eiles_2017}.  
The results are summarized in Table \ref{Tab} and demonstrate an excellent agreement with the measured dipole moments.

Finally, we measured the lifetime of the butterfly molecules by reducing the pulse duration to \SI{1}{\micro s} and varying the time of ionization. We only count Rb$^+$ ions with zero momentum as $\ell$-changing collisions leads to ions with large momentum. The ion counts are then fitted with an exponential function as shown in Fig \ref{fig:lifetime}. The vibrational ground state shows a longer lifetime than the corresponding atomic resonance. This is a signature of the high-$\ell$ admixture of the molecular state, as the lifetime increases with angular momentum. This was also observed in previous trilobite studies \cite{Althoen2023}. Additionally, we find that the $\nu = 1$ vibrational state has a significantly shorter lifetime than does the ground state. This can be attributed to its energetic proximity to
the avoided crossings between the $M_L=0$ and $M_L=\pm1$ curves, leading to increased non-adiabatic decay to smaller internuclear distances, consequently resulting in $\ell$-changing collisions or associative ionization. However, the computed lifetimes of these states are substantially longer than those measured, suggesting that effects outside the scope of this calculation, such as three-body collisions or additional potential curve coupling to higher partial wave states, may play a role in reducing the lifetime. 

The work presented here constitutes the first observation of a Rydberg "butterfly" molecule where the Rydberg electron and the valence electron of the ground-state atom are in a singlet ($S=0$) configuration. From comparison of the binding energies, dipole moments, and lifetimes of these molecular states with theory, we have confirmed the character of these molecular states and verified the accuracy of the theoretical description based on the energy-dependent scattering phase shifts for the $^1P_1$ channel calculated in Refs.~\cite{Khuskivadze_2002,Eiles_2018}. The unassigned peaks visible in the spectrum point towards the importance of including higher-order partial wave scattering in the theoretical description and, given the smaller internuclear distances and low principal quantum number at play in this experiment, possible improvements to the underlying Fermi model.  The successful preparation of such singlet character-dominated molecular states is a key stepping stone towards the production of heavy Rydberg ion-pair states.

Dissociation of such ion-pair states will lead to the creation of ultracold Rb anions, which have not been observed to date in ultracold atomic traps; and potentially to the formation of strongly coupled two-component plasmas~\cite{murillo}.
\newline

We would like to thank Max Althön for helpful discussions. This project is funded by the German science foundation DFG, project numbers 460443971, 316211972 and INST 248/268-1.  H.R.S. acknowledges funding at ITAMP through the NSF.
P.S. acknowledges support by the cluster of excellence CUI: Advanced Imaging of Matter, of the Deutsche Forschungsgemeinschaft (DFG) EXC-2056 project ID - 309715994.
\newline

\bibliographystyle{apsrev4-2}
\bibliography{references}

\vspace{-2mm}

\end{document}